# The way we cite: common metadata used across disciplines for defining bibliographic references

Erika Alves dos Santos[1][0000-0002-3162-767X], Silvio Peroni[2,3][0000-0003-0530-4305] and Marcos Luiz Mucheroni[4][0000-0002-5273-7392]

[1] Fundação Jorge Duprat Figueiredo de Segurança e Medicina do Trabalho (Fundacentro), São Paulo, Brazil
`erika.santos@fundacentro.gov.br`
[2] Research Centre for Open Scholarly Metadata, Department of Classical Philology and Italian Studies, University of Bologna, Bologna, Italy
`silvio.peroni@unibo.it`
[3] Digital Humanities Advanced Research Centre (/DH.arc), Department of Classical Philology and Italian Studies, University of Bologna, Bologna, Italy
[4] School of Communication and Arts (ECA), Department of Information & Culture (CBD), University of São Paulo, São Paulo, Brazil
`mucheroni.marcos@gmail.com`

**Abstract.** Current citation practices observed in articles are very noisy, confusing, and not standardised, making identifying the cited works problematic for humans and any reference extraction software. In this work, we want to investigate such citation practices for referencing different types of entities and, in particular, to understand the most used metadata in bibliographic references. We identified 36 types of cited entities (the most cited ones were articles, books, and proceeding papers) within the 34,140 bibliographic references extracted from a vast set of journal articles on 27 different subject areas. The analysis of such bibliographic references, grouped by the particular type of cited entities, enabled us to highlight the most used metadata for defining bibliographic references across the subject areas. However, we also noticed that, in some cases, bibliographic references did not provide the essential elements to identify the work they refer to easily.

**Keywords:** bibliographic references, citations, publication metadata, publication types, citation behaviours.

## 1 Introduction

Citations are fundamental tools to track how science evolves over time [1]. Indeed, citation networks are the instruments that link scientific thinking, forming a complex chain of documents related to each other that enables the highlighting of research trends within the various scholarly disciplines. The creation of such a network is possible thanks to the effort that authors and publishers invest in preparing particular elements of their articles: the bibliographic references. Bibliographic references are the means for creating conceptual citation links between a citing and cited entities and carry an important function: providing enough metadata to facilitate an agent (whether a human or a machine) to identify the cited works. Thus, providing precise bibliographic metadata of cited works is crucial for enabling citation networks to satisfactorily and efficiently contribute to the intellectual exchange among researchers.

Despite the massive number of citation style manuals released in the past years that have had the goal of providing standardised approaches to the definition of bibliographic references (and, in particular, their metadata), some prior studies, such as [12] and [7], have shown how the current citation practices are very noisy, confusing, and not standardised at all. For instance, several disciplinary journals often avoid adopting standardised citation style manuals and define their own (yet another) citation style [3]. This considerable heterogeneity in the adoption of citation guidelines, combined with the variability of the types of cited works (and, thus, of related metadata) that may include articles, datasets, software, images, green literature, etc., makes the identification of the cited works problematic for humans and also (and in particular) for any reference extraction software used for building bibliographic metadata repositories and citation indexes.

In this work, following prior studies we run on similar topics [7-8], we want to investigate existing citation practices by analysing a huge set of articles to measure which metadata are used across the various scholarly disciplines, independently from the particular citation style adopted, for defining bibliographic references. In particular, we want to answer the following research questions (RQ1-RQ3):

1. Which entities are cited by articles published in journals of different disciplines?
2. What is the standard metadata set used across such disciplines for describing cited works within bibliographic references?
3. Is there any mechanism in place (i.e. hypertextual links) to facilitate the algorithmic recognition of where a bibliographic reference is cited in the text?

The rest of the paper is organised as follows. In Section 2, we introduce some related works concerning our research. In Section 3, we present the material and methods we have used for performing our analysis. Section 4 introduces the results of our analysis, which are discussed in Section 5. Finally, in Section 6, we conclude the paper by sketching out some future works.

## 2     Related works

In the past, several works addressed studies and analyses of bibliographic references from different perspectives. One of the essential works in the area is authored by Sweetland [12]. In his work, he highlighted the functions conveyed by bibliographic references and citation style manuals and the errors in the reference lists and in-text citations that represent a crucial issue for accomplishing such functions. In particular, he identified the use of a great variety of formats for referencing cited articles that increased the chances of misunderstanding referencing guidelines proposed by the journals, which, consequently, contributes to the high errors in bibliographic metadata description. A recent study we performed [7], run against a larger corpus of journal articles and bibliographic references and used as starting point of the work presented in this paper, confirmed that many of the concerns highlighted by Sweetland are in place still today, thus showing that the situation has not changed in the past 32 years.

Some mistakes identified in bibliographic references may be conveyed by limited clarity in describing particular publication types cited in articles. Indeed, depending on the type of the cited works, metadata of bibliographic references may change a lot: from *author(s)*, *year of publication*, *article title*, *journal name*, *volume*, *issue*, *page numbers*, typical of journal articles, to *author(s)*, *year of publication*, *article title*, *complete title proper of proceedings volume in which it occurs*, *statements of responsibility for the proceedings*, *series statement*, *place*, *publisher* and *page numbers*, typical of conferences [11]. However, sometimes, journal citation styles fail to address all the possible publication types cited by the authors of a citing article [7].

One study introduced by Heneberg [2], among the most relevant ones focussing on the analysis of specific disciplines, analysed the percentage of uncited publications that were not journal original research articles or reviews authored by scientists in Mathematics, Physiology and Medicine who either received Fields medals or were Nobel laureates. He discovered that the most significant part of these uncited publications listed in Web of Science (WoS) was mainly editorial material, progress reports (e.g. abstracts presented at conferences), and discussion-related publications (e.g. letters to the editor). Only a small number of research articles and reviews in journals were left uncited, thus highlighting how the types of the publications seemed to be a relevant characteristic which explained, at least to a certain extent, why part of the works of even influential authors are not cited at all.

In another work, Kratochvíl *et al.* [3] analysed the declared referencing practices of 1,100 journals in the biomedical domain. They discovered that, even if there exist still today several citation guidelines for biomedical research, a considerable number of biomedical journals preferred to adopt their own style and that the most essential metadata used when referencing cited works were *author(s)*, *cited work title*, and *year of publication*. However, helpful metadata (e.g. *DOI*), recognised from the answers to more than 100 surveys the authors performed, were not included in several of the citation styles adopted by the journals in the corpus analysed.

Other studies have concerned the analysis of citations to specific kinds of publications, e.g. data (in a broader sense, i.e. including datasets and software). For instance, Park *et al.* [5] analysed hundreds of biomedical journals to measure the number of *formal* citations to data (i.e. specified by including a bibliographic reference describing them) against the *informal* citations (i.e. mentions contained within the text of an article, e.g. by simply adding their URL). They highlighted how informal citations to data were the most adopted approach due mainly to the absence of explicit requirements by the publisher to correctly add them as bibliographic references (showing an inadequate citation type coverage in the citation styles adopted) and, in part, to the limited familiarity of the authors when dealing with formal citations to data. Indeed, several studies, such as [4], stressed that mastering citation styles is a complex activity and that there is a need to reflect on (and even redesign) citation styles to address current citation habits.

## 3     Materials and methods

The articles from which we have extracted the bibliographic references to analyse for this study were obtained from a selection of the journals included in the SCImago Journal & Country Rank (https://www.scimagojr.com/). Following a methodology we defined, which is introduced with more details in [10] and that has been already successfully adopted in previous studies [7-8], we first selected the most cited journals in each of the 27 subject areas listed in SCImago in the 2015-2017 triennium according to the SCImago total cites ranking. We grouped these subject areas in five macro categories: *Health Sciences* [H] (including the subject areas *Medicine* [S1], *Nursing* [S2], *Veterinary* [S3], *Dentistry* [S4], *Health Professions* [S5]), *Social Sciences and Humanities* [S] (including *Arts and Humanities* [S6], *Business, Management and Accounting* [S7], *Decision Sciences* [S8], *Economics, Econometrics and Finance* [S9], *Psychology* [S10] and *Social Sciences* [S11]), *Life Sciences* [L] (including *Agricultural and Biological Sciences* [S12], *Biochemistry, Genetics and Molecular Biology* [S13], *Immunology and Microbiology* [S14], *Neuroscience* [S15], *Pharmacology, Toxicology and Pharmaceutics* [S16]), *Physical Sciences* [P] (including *Chemical Engineering* [S17], *Chemistry* [S18], *Computer Science* [S19], *Earth and Planetary Sciences* [S20], *Energy* [S21], *Engineering* [S22], *Environmental Science* [S23], *Materials Science* [S24], *Mathematics* [S25], and *Physics and Astronomy* [S26]), and *Multidisciplinary* [M] (including the subject area *Multidisciplinary* [S27] mainly involving big magazine and journals). The sample we obtained was the proportional representation of each subject area at SCImago Ranking in terms of dimension. We included only one journal from each publisher under the same subject area to avoid having, under the same subject area, journals sharing similar editorial policies.

Each journal in the sample was represented by five articles (in PDF format) published in the most recent issue (excluding special issues that sometimes adopt diversified journal policies for referencing) published between October 1$^{st}$ and October 31$^{st}$, 2019. For journals not releasing any issue in this period, the sample considered the immediately previous issue published before October 1$^{st}$. For issues containing more than five articles, the selection adopted a probabilistic systematic random sampling technique based on the average number of articles published by the journal in the period mentioned above. As for the journals containing less than five articles, the sample considered all those attending the selection criteria described in detail in [10].

Starting from such sample, we considered the total of 34,140 bibliographic references composing the bibliographic references lists of the selected 729 articles (172 in Health Sciences, 191 Social Sciences and Humanities, 114 in Life Sciences, 232 in Physical Sciences, and 20 in Multidisciplinary) which were analysed to detect the types of the cited works in each discipline and the structure of bibliographic references for each type of cited work, considering different reference styles' formatting guidelines. In particular, we identified the descriptive elements (introduced in Table 1) adopted for the bibliographic references for each type of cited work.

**Table 1.** Kinds of metadata retrieved in the bibliographic references analysed.

| Title | Series statement |
|---|---|
| 1. Chapter title | 34. Conference date |
| 2. Chapter title in English (when original title is in another language) | **Identifier for manifestation** |
| 3. Conferences' title | 35. Abstract number |
| 4. Journals' title (abridged format) | 36. Article ID within publisher's webpage |
| 5. Journals's title (full format) | 37. Article number part note |
| 6. Journals's title in English (for titles in other languages) | 38. Chapter number |
| | 39. ISBN number |
| 7. Newspaper/magazine title | 40. Paper number |
| 8. Proceedings' title | 41. Patent number |
| 9. Session title | 42. Technical report number |
| 10. Works' subtitle in original language | 43. Work number |
| 11. Works' title in original language | 44. Work number within the conference |
| 12. Works's title in English (when original title is in another language) | 45. Working paper number |
| | **Carrier type** |
| **Statement of responsibility** | 46. Content type / media type / carrier type (general material designation in AACR2) |
| 13. Author full name | |
| 14. Chapter author | **Extent** |
| 15. Proceedings' editor | 47. Abridged work pagination length (e.g. 80-9) |
| 16. Translator | 48. Mentioned excerpts pages range (e.g. 80-89) |
| 17. Work's author or editor | 49. Work's first page number (e.g. 80) |
| | 50. Work full pagination length (e.g. 80-89) |
| **Edition statement** | 51. Work's total number of pages (e.g. 80 p.) |
| 18. Edition number | |
| 19. Issue number | **General notes** |
| 20. Revision number | 52. Work's language note |
| 21. Version number | 53. Supplemental issue note |
| 22. Volume number | 54. Special issue note |
| | 55. Supplementary content note |
| **Numbering of serials** | 56. General notes |
| 23. Series number | 57. Unpublished note |
| | 58. In press note |
| **Publication statement** | 59. Database system number |
| 24. Conference date | |
| 25. Conference place | **Online availability notes** |
| 26. Date of citation (date of access) | 60. Hypertext hyperlink (URL) |
| 27. Date of last update/revision | 61. DOI string or DOI URL |
| 28. Day of publication | 62. Online availability note |
| 29. Month of publication | 63. Institutional link (university department) |
| 30. Place of publication | |
| 31. Proceedings date of publication | **Miscellaneous** |
| 32. Publisher (or granting institutions for thesis and dissertations) | 64. Latin expression "in" (i.e. for book chapters or conference papers in a proceedings) |
| 33. Year or date of publication | |

Such descriptive elements were classified according to the Resource Description & Access (RDA) core elements (https://www.librarianshipstudies.com/2016/03/rda-core-elements.html). In addition, we also analysed all the in-text reference pointers – e.g. "(Doe et al., 2022)" and "[3]" – denoting all the bibliographic references in our sample to see how many of them are accompanied by a link pointing to the related bibliographic reference they denote.

## 4      Results

All the data gathered in our analysis are available in [9]. In the first stage of the analysis we considered all the 34,140 bibliographic references composing our sample, that we used to identify the following different kinds of publications (RQ1): articles [a], books and related chapters [b], manuscripts [c], technical reports and related chapters [d], webpages [e], proceeding papers [f], conference papers [g], grey literature [h], data sheets [i], forthcoming chapters [j], forthcoming articles [k], unpublished material [l], standards [m], working papers and preprints [n], e-books and related chapters [o], newspapers [p], online databases [q], web videos [r], patents [s], software [t], manuals/guides/toolkits [u], personal communications [v], book series [w], other kinds of publications (including memorandum, governmental official publications, legislation, informative materials, audio records, motion pictures, speeches, photographs, slide presentation, podcasts, engravings, lithography and television shows) [y], and unrecognised publications [z].

As summarised in Table 2, articles, books (and their chapters), and proceeding papers were the first, second and third most cited types of publications across all the subject areas. The same seven types of publications corresponded to at least 50% of the total bibliographic references in each subject area, namely articles (83.55%), books and their chapters (7.93%), proceeding papers (2.53%), webpages (1.30%), technical reports (1.17%), working papers and preprints (0.67%) and conference papers (0.51%). However, these types did not comprise some other publication kinds cited by specific disciplines. For instance, grey literature is the eighth most cited type of work across all subject areas (0.47% of total bibliographic references). Still, it is the third most cited type of publication in arts and humanities articles (S6) and the fourth most cited type in chemical engineering (S17), decision sciences (S8) and mathematics (S25) articles. Thus, considering only the most cited types of publications overall does not properly represent the actual citing habits across the subject areas since some subject areas (e.g. social sciences – S11) tend to cite a greater variety of types of publications while others (e.g. dentistry – S4) only a few types. In addition, as highlighted in Table 2, the types of publications supporting discussions across subject areas may vary.

To understand the variability of the metadata for defining bibliographic references across the macro areas, we decided to select the seven most cited types of publications in each subject area to assure that the analysis coverage includes the most cited types of publications from the subject areas' perspective. After this selection, all the types of publications in Table 1 were considered except manuscripts (c), forthcoming chapters (j), web videos (r), other kinds (y) and unidentified types of publications (z).

**Table 2.** All the different kinds of publications (a-z, as defined in the text) cited by the various subject areas (column S, S1-S27 as defined in Section 3) grouped in macro areas (column A, values as defined in Section 3). The colours of the squares represent the proportion of citations from the citing articles of S1-S27 to the a-z publication kinds, described as follows: ■ >80%, ■ >60%, ■ >40%, ■ >20%, ■ >10%, ■ >5%, ■ >1%, ■ >0%

| A | S | Kinds of cited publications |
|---|---|---|
| | | a b c d e f g h i j k l m n o p q r s t u v w y z |
| H | S1 | |
| | S2 | |
| | S3 | |
| | S4 | |
| | S5 | |
| S | S6 | |
| | S7 | |
| | S8 | |
| | S9 | |
| | S10 | |
| | S11 | |
| L | S12 | |
| | S13 | |
| | S14 | |
| | S15 | |
| | S16 | |
| P | S17 | |
| | S18 | |
| | S19 | |
| | S20 | |
| | S21 | |
| | S22 | |
| | S23 | |
| | S24 | |
| | S25 | |
| | S26 | |
| M | S27 | |

The 33,786 bibliographic references concerning such most significant types of publications were individually analysed to identify their descriptive elements (i.e. metadata) according to those introduced in Table 1. We have tracked all the bibliographic elements appearing in the bibliographic references of our sample, and we marked all the elements specified in at least one bibliographic reference of at least 50% of the articles composing each subject category. Finally, we have computed the most used descriptive elements for each type of publication mentioned above by considering each macro area's most used descriptive elements. In practice, a descriptive element was selected if it was one of the most used in all the macro areas. The result of this analysis is summarised in Table 3 (RQ2).

In the last part of our analysis, we identified if the in-text reference pointers – e.g. "(Doe et al., 2022)" or "[3]" – included in all the articles of our sample are hypertextually linked to the respective bibliographic references they denote (RQ3). The result of such analysis is shown in Fig. 1.

**Table 3.** Most used metadata in bibliographic references – the numbers identify the kinds of metadata as introduced in Table 1. H: Health Sciences, S: Social Sciences and Humanities, L: Life Sciences, P: Physical Sciences, M: Multidisciplinary, A: average.

| Articles | Books | Book chapters |
|---|---|---|
| H  4,11,17,22,33,50 | H  11,17,18,30,32,33 | H  1,11,14,17,30,32,33,48,64 |
| S  5,11,17,19,22,33,50 | S  11,17,30,32,33 | S  1,11,14,17,30,32,33,48,64 |
| L  4,11,17,22,33,36,50 | L  11,17,30,32,33 | L  1,11,14,17,30,32,33,48,64 |
| P  4,11,17,22,33,50 | P  11,17,30,32,33 | P  1,11,14,17,30,32,33,48,64 |
| M  4,11,17,22,33,36 50 | M  11,17,32,33 | M  1,11,14,17,30,32,33,38,48,64 |
| **A  11,17,22,33,50** | **A  11,17,32,33** | **A  1,11,14,17,30,32,33,48,64** |

| Technical reports | Webpages | Proceeding papers |
|---|---|---|
| H  4,11,17,22,33,50 | H  11,17,26,60 | H  3,11,17,33,48,64 |
| S  5,11,17,19,22,33,50 | S  11,17,33,60 | S  8,11,17,30,32,33,48,64 |
| L  4,11,17,22,33,36,50 | L  11,17,33,60 | L  8,11,17,32,33,48,64 |
| P  4,11,17,22,33,50 | P  11,17,33,60 | P  8,11,17,32,33,48,64 |
| M  4,11,17,22,33,36,50 | M  11,17,60 | M  3,8,11,17,32,33,48,64 |
| **A  11,17,22,33,50** | **A  11,17,60** | **A  11,17,33,48,64** |

| Conference papers | Grey literature | Data sheets | Technical rep. chapters |
|---|---|---|---|
| H  3,11,17,33 | H  11,17,30,32,33,46 | H  No citations | H  1,11,14,22,30,32,33,60 |
| S  3,11,17,25,33 | S  11,17,32,33,46 | S  No citations | S  1,11,14,30,32,33 |
| L  3,11,17,25,33 | L  11,17,30,32,33,46 | L  No citations | L  No citations |
| P  3,11,17,25,33 | P  11,17,30,32,33,46 | P  11,32,33 | P  No citations |
| M  No citations | M  No citations | M  No citations | M  No citations |
| **A  3,11,17,33** | **A  11,17,32,33,46** | **A  11,32,33** | **A  1,11,14,30,32,33** |

| Forthcoming articles | Unpublished | Standards | Working papers |
|---|---|---|---|
| H  4,11,17,33,58,61 | H  No citations | H  No citations | H  11,17,26,30,32,33,60 |
| S  5,11,17,58 | S  11,17,33,57 | S  11,17,33 | S  11,17,33,45,60 |
| L  4,11,17,22,29,33,58,60 | L  No citations | L  11,17,30,33 | L  11,17,26,33,61 |
| P  4,11,17,33,58 | P  11,17,32,33,57 | P  11,17,18,33,51 | P  11,17,33,60 |
| M  No citations | M  No citations | M  No citations | M  11,17,32,33,60 |
| **A  11,17,58** | **A  11,17,33,57** | **A  11,17,33** | **A  11,17,33** |

| E-books | Newspapers | Online databases |
|---|---|---|
| H  11,17,30,32,33 | H  No citations | H  11,17,21,26,33,60 |
| S  11,17,30,32,33 | S  7,11,17,28,33,60 | S  11,17,32,33,60 |
| L  11,17,26,30,32,33,60 | L  No citations | L  11,17,21,33 |
| P  11,17,18,26,33,39,60,61 | P  No citations | P  11,17,21,32,33,46,60,61 |
| M  No citations | M  No citations | M  No citations |
| **A  11,17,33** | **A  7,11,17,28,33,60** | **A  11,17,33** |

| E-books chapters | Patents | Software |
|---|---|---|
| H  1,11,14,17,30,32,33,48 | H  11,17,33,41 | H  11,17,30,32,33 |
| S  1,11,17,30,32,33,64 | S  No citations | S  11,17,30,32,33,46 |
| L  No citations | L  11,17,33,41,46 | L  11,17,21,26,30,32,33,60 |
| P  1,11,14,17,26,33,60,64 | P  11,17,30,33,21,41,48 | P  11,17,21,33,60 |
| M  No citations | M  11,17,33,41,60 | M  No citations |
| **A  1,11,17,33** | **A  11,17,33,41** | **A  11,17,33** |

| Manual/guides/toolkits | Personal communications | Book series |
|---|---|---|
| H  11,17,30,32,33,60 | H  11,17,30,32,33,60 | H  1,14,19,22,32,33,34,47,49,61 |
| S  11,17,30,32,33 | S  11,17,28,33,46,60 | S  No citations |
| L  11,17,32,33 | L  No citations | L  No citations |
| P  11,17,21,32,33 | P  No citations | P  1,14,19,22,32,33,34,49,61 |
| M  No citations | M  No citations | M  No citations |
| **A  11,17,32,33** | **A  11,17 33,60** | **A  1,14,19,22,32,33,34,49,61** |

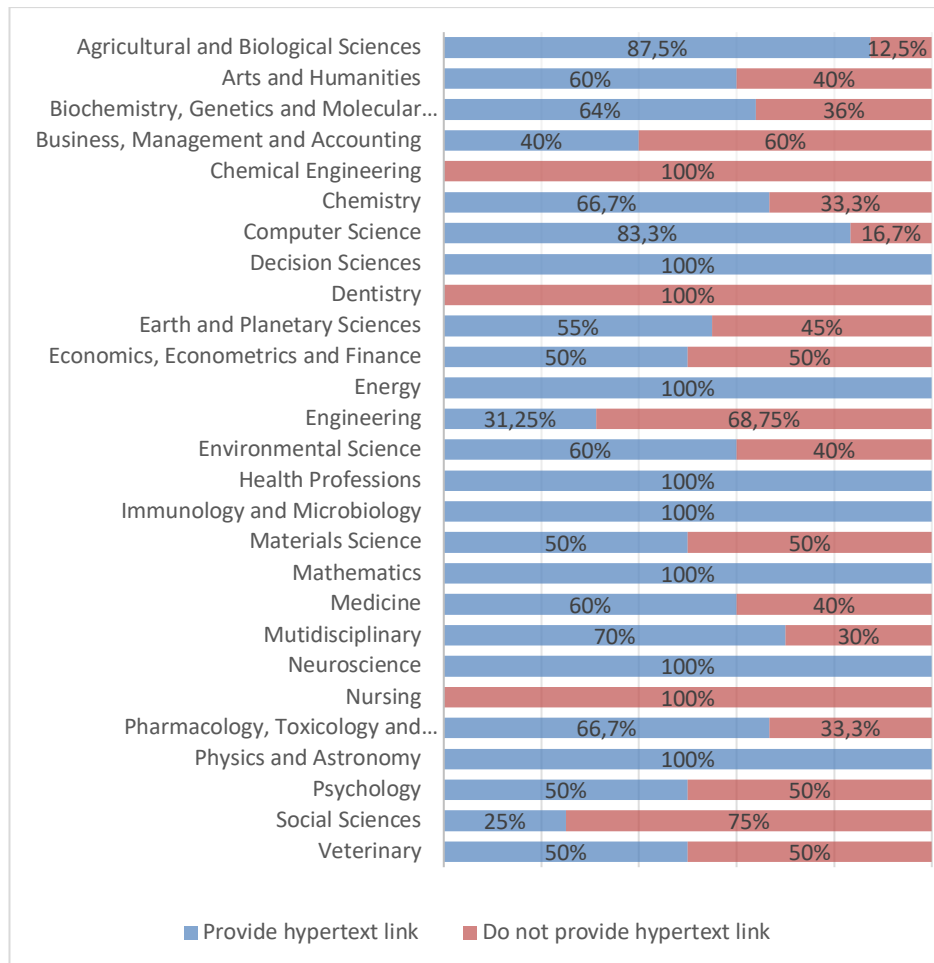

**Fig. 1.** Distribution of articles per subject area providing in-text reference pointers – e.g. "(Doe et al., 2022)" and "[3]" – hypertextually linked to the bibliographic references denoted.

## 5   Discussion and lessons learnt

The data in Table 2 suggest that articles are the most used channel to communicate scientific findings across all the subject areas. However, books were observed among the three most cited types of publications in all disciplines considered in our sample. In addition, we observed considerable variability in the types of publications cited by the articles composing our sample – we found 36 different types of publications within disciplines. Such variety suggests and reveals some citing habits across disciplines. For instance, we noticed a considerable portion of bibliographic references for which we could not identify which type of publication is referred to (columns "z" in Table 2), considering the data provided in the bibliographic references. This suggests that either

1010reference styles adopted by the journal were unclear or did not provide enough instructions on describing certain types of publications. We could also speculate that part of these issues derived from the lack of attention that authors and publishers sometimes put when writing/revising bibliographic references; however, this aspect should be investigated in more detail.

Still looking at the results in Table 2, it seemed that some disciplines, e.g. the humanities and social sciences, cited many publication types. This suggests that the discussions on such disciplines demand more comprehensive approaches. Second, reference styles adopted by such disciplines should provide more extensive guidelines for describing citing and referencing data, i.e. they should provide instructions on describing a greater variety of publications. The lack of specific guidelines for describing uncommon types of publications across disciplines, such as lithographs and engravings (which appeared in some social sciences articles), contributes to the number of unidentifiable bibliographic references mentioned above.

Despite the existence of thousands of reference styles and standards to guide the use and interpretation of bibliographic metadata uniformly, we observed (Table 3) that the representation of the information is approached differently across subject areas and, in general, macro areas: the same type of publication may have different descriptions in different disciplines. This may suggest a failure of reference styles' purposes concerning their role in standardising bibliographic references on a large scale.

For instance, among their various purposes, bibliographic references act like sources of information and, from this perspective, the efforts to provide (at least) the necessary metadata for the proper identification of the referred publications are worthwhile and essential as a means for retrieving the cited works in external sources, such as bibliographic catalogues and bibliographic databases. However, we noticed that such kinds of metadata were not always provided. Even if the title of the cited works (11 in Table 3) is one of the most used metadata across all the macro areas, we observed that bibliographic references in some articles did not always provide it. For instance, in 27% of articles from Physical Sciences, we noticed that bibliographic references pointing to web pages did not provide the title of the cited work. At the same time, they include a URL or a persistent identifier (e.g. DOI) to enable accessing the cited publication. Indeed, in some cases, the article's title itself is not a mandatory element for allowing its retrieval (e.g. if a DOI is present). However, it is a crucial element to correctly identify the cited work, which is one of the primary purposes of bibliographic references. Similarly, considering bibliographic references referring to articles, we observed that metadata like the issue number (19 in Table 3) in which the cited article was published were omitted in most macro areas.

Another point highlighted in Table 3 is that different publication types may have different characteristics. Indeed, the description of different types of publications may demand different types of metadata, which do not necessarily play the same role in the identification of the cited work and, because of that, may have different levels of importance in terms of facilitating the task of identifying the cited work and such issues should be considered by metadata treatment tools, like the ontologies.

We also noticed that part of the bibliographic references providing URLs to the cited works did not provide the date in which that content was consulted. This may represent issues in later retrieving of such content because, unlike press sources of information that usually are modified after their release, online sources are susceptible to amendments and might even become unavailable at any time without prior notification.

In general, concerning the uniformity of the metadata provided by bibliographic references referring to specific types of publications overall, we can notice that, in most cases, there is a relative (i.e. poor) uniformity. Indeed, the metadata referring to the same type of publications vary across disciplines.

A careful analysis of the data in Table 3 showed other deficits in normalising bibliographic references, even when they reference the same type of publication. For instance, considering those referring to articles, we noticed that 71.59% provide the title of the journal which has published the cited article in the abridged format. This may be a problem for identifying the full title of a journal since, even if there exist several sources defining journals titles abbreviations such as the NLM Catalog (https://www.ncbi.nlm.nih.gov/nlmcatalog/journals/) and the CAS Source Index (CASSI, https://cassi.cas.org), the big issue is that the abbreviation for a particular journal title may be different considering different sources guidelines. This may have negative consequences for the precise identification of the referred journal and, consequently, for its retrieval. Thus, to ensure the correct interpretation (also in the context of computational approaches), the journal title abbreviation should be accompanied by the indication of the source on which it was based.

By analysing the most used metadata (rows "A" in Table 3), we can observe that bibliographic references usually dismiss important elements that help readers identify the cited works. For instance, the DOI is not included in the most used metadata in the bibliographic references referring to articles, as the ISBN is not part of the most used metadata in the bibliographic references referring to books and book chapters.

Overall, the most used metadata gathered are enough, in general, to identify the publications the bibliographic references refer to. We did notice some peculiar situations, however. The most used metadata for proceedings do not comprise the title of the proceedings in which the cited work was published nor the title of the conference in which the cited work was presented, even if these kinds of metadata were indeed used in the macro areas: Health Sciences and Multidisciplinary included the title of the conference (3), while Social Sciences, Life Sciences, Physical Sciences and, again, Multidisciplinary included the title of the proceedings (8).

For some publication types – i.e. software (t); manual, guides and toolkits (u), data sheets (i), standards (m) and personal communications (v) – we noted that bibliographic references do not provide any information concerning the nature of the document (i.e. the "general material designation in AACR2", carrier type, point 46, in Table 1). The description of less-traditional types of publications – i.e. those except articles, books, and other similar textual publications – requires a clear indication of the type of publication being cited for allowing its immediate identification from the metadata provided in bibliographic references. For instance, grey literature usually provides a short note like "master thesis" or "doctoral thesis", which enables the reader to understand the format of the cited work immediately.

The links between in-text reference pointers and the bibliographic references they denote (RQ3) are helpful tools to formalise the connections between the text of the citing article (i.e. the sentences including the in-text reference pointers, the related paragraphs and sections) and the correspondent cited works referenced by the bibliographic references. Around 49% of the articles in our sample provide such a feature (Fig. 1). Indeed, having such mechanisms in place simplifies, in principle, the development of computational tools to track where cited works are referred to in the text of the citing articles, thus facilitate the computational recognition of citation sentences [6] and, by analysing these, of citation functions [13], i.e. the reason an author cites a cited work – because it reuses a method defined in the cited work, because it either agrees or disagrees with concepts and ideas introduced in the cited works, etc.

However, 51% of the articles did not specify such links, which is a barrier to identifying the position where a citation is defined in the text. Of course, one could use natural language processing tools and other techniques to retrieve the in-text reference pointers referring to bibliographic references in the text, but this is made complex by the heterogeneity of the formats used to write bibliographic references and in-text reference pointers, as highlighted in [7].

Finally, it is worth mentioning that our analysis is not free from limitations. For instance, we considered only one type of publication for the citing entities, i.e. journal articles. Indeed, since they represent the main types of publication cited across all the subject areas, at least according to our analysis, they should be a reasonable sample acting as a proxy of the entire population of the citing publications in all the subject areas considered. However, it would be possible that different publication types of citing articles may convey different citation behaviours. We leave this analysis to future studies.

## 6 Conclusions

This work has focussed on presenting the results of an analysis of 34,140 bibliographic references in articles of different subject areas to understand the citing habits across disciplines and identify the most used metadata in bibliographic references depending on the particular type of cited entities. In our analysis, we observed that the bibliographic references in our sample referenced 36 different types of cited works. Such a considerable variety of publications revealed the existence of particular citing behaviours in scientific articles that varied from subject area to subject area. In the future, further investigation should be performed to understand, for instance, why software was not listed among the most cited type of works in Computer Science while being one of the main topics discussed in several areas of Computer Science research.

**Acknowledgements.** This work was partly conducted when Erika Alves dos Santos was visiting the Department of Classical Philology and Italian Studies, University of Bologna. The work of Silvio Peroni has been partially funded by the European Union's Horizon 2020 research and innovation program under grant agreement No 101017452 (OpenAIRE-Nexus).